\documentclass[showkeys]{revtex4-1}

\usepackage{graphics}

\begin{document}

\newcommand{\bmb}[1]{\mathbf{#1}}
\newcommand{\ms}{m^{*}}
\renewcommand{\S}{{\sf S}}
\renewcommand{\dag}{+}
\newcommand{\bbm}{\bmb}
\title{Undamped plasmon-polariton propagation along metallic nano-chain including nonlinear effects}

\author{Witold Jacak         }

\affiliation{Institute of Physics, Wroc{\l}aw University of Technology, Wyb. Wyspia\'nskiego 27, 50-370 Wroc{\l}aw, Poland \\
              \email{witold.aleksander.jacak@pwr.wroc.pl}  }

\date{Received: date / Accepted: date}

\begin{abstract}
The nonlinear theory of collective plasmon-polariton propagation along the infinite chain of metallic nanoparticles is developed within RPA quasiclassical approach to surface plasmons in  large nano-spheres ($10-50$ nm for radius) of Au or Ag. The wave type self-modes in the chain are determined and analyzed with respect to nano-sphere  size and chain-separation parameters. At some regions for parameters  the  undamped modes occur. They were found  on the rim of stability within the linear theory. The nonlinear corrections  stabilize, however, diverging modes of the linear approach and considerably enlarge the region of undamped propagation.  Nonlinearity is associated with relativistic corrections to the Lorentz friction. According to the nonlinear behavior, the region of parameters when undamped  stable modes occur, covers the instability region of the linear theory. The remarkable property of fixed amplitude of the undamped collective propagating wave independently of initial conditions (even if they are extremely small) has been identified and described. The group velocity of this wave propagation is calculated.

\end{abstract}

\maketitle

\section{Introduction}
\label{intro}

Experimental and theoretical investigations of plasmon excitations in metallic nano-crystals rapidly grew up mainly due to  possible applications in photo-voltaics and microelectronics. A significant enhancement of absorption of  incident light in photodiode-systems with  active surfaces covered with nano-size metallic particles (of Au, Ag or Cu) with planar density $10^8$-$10^{10}$/cm$^2$ was observed \cite{wzr3a,wzmocn1,wzr2,konk,wzmocn2,mof}. These findings are of practical importance for enhancement of solar cell efficiency, especially for developing of thin film cell technology. On the other hand, hybridized states of  surface plasmons and photons result in plasmon-polaritons \cite{zastos,maradudin}, which are of high importance for applications in photonics and microelectronics \cite{zastos,deabajo}, in particular,  for sub-diffraction transportation of converted light energy and information in metallically modified structures in nano-scale \cite{maradudin,atwater1}.

Surface plasmons in nano-particles have been widely investigated since their classical description by Mie \cite{Mie}. Many particular studies, including numerical modelling of multi-electron clusters, have been carried out \cite{brack,brack1}. They were mostly developments of Kohn-Sham attitude in form of LDA (Local Density Approximation) or TDLDA (Time Dependent LDA) for small metallic clusters only \cite{brack,brack1,ekardt,ekardt2,kresin}, up to ca 200 electrons (limited for larger clusters by  numerical calculation constraints that grow rapidly with the number of electrons). The random phase approximation (RPA) was formulated \cite{rpa} for description of volume plasmons in bulk metals  and utilized also for confined geometry mainly in a numerical  or semi-numerical manner \cite{brack,brack1,kresin}. Usually, in these analyses the jellium model was assumed for description of positive ion background in the metal and the dynamics was addressed to the electron system only \cite{brack,ekardt,kresin}. Such a model is preferable for clusters of simple metals, including noble metals (also transition and alkali metals).

In the present paper we apply the RPA description using a semiclassical approach for a large metallic nano-sphere (with radius of several tens nm, and with $10^5$-$10^7$ electrons), in an all-analytical calculus version \cite{jacak5}. The electron liquid oscillations of compressional and translational type result in excitations  inside the sphere and on its surface, respectively. They  are  referred to as volume and surface plasmons. Damping of plasmons due to electron scattering and due to radiation losses (accounted for via the Lorentz friction force) is included. The shift of the resonance frequency of dipole-type surface plasmons (only such plasmons are induced by homogeneous time-dependent electric field), due to damping phenomena, well fits with the experimental data for various nano-sphere radii \cite{jacak11}.

Collective dipole-type surface plasmon oscillations in the linear chain of metallic nano-spheres were then analyzed and wave-type plasmon propagation along the chain was described \cite{jacak10,maradudin}. A coupling in the near field regime between oscillating dipoles in neighboring nano-spheres, together with retardation effects for energy irradiation, allowed for appearance of undamped propagation of plasmon waves (called plasmon-polaritons) along the chain in the experimentally realistic region of values  of the separation of spheres in the chain and of the nano-sphere radii. This effect is of a particular significance for  plasmon arranged non-dissipative and sub-diffraction transport of light converted energy and information along metallic chains for possible applications in nano-electronics.

The undamped mode of plasmon-polaritons occurs, however, on the rim of stability of the linear approach. The zero damping rate separates the region with positive its value (corresponding to ordinary attenuation of plasmon-polaritons) and the region with negative damping rate (corresponding to unstable  modes). The latter exhibits unphysical behavior being the artefact of the linear approximation. In order to regularize the description, the  nonlinear corrections must be thus included. The nonlinear corrections may be associated with the Lorentz friction forces. Small relativistic contribution to this friction has nonlinear character and quenches instable divergent modes.  In the result, the instability region of linear approach is entirely covered by the region of undamped wave propagation with the amplitude accommodated, however, to nonlinearity scale and independent of the initial condition (despite of its magnitude). This phenomenon, familiar in other nonlinear systems \cite{mit}, seems to be of a particular significance for understanding of collective plasmon excitations  with interesting possible applications. 

\section{Damping of plasmons in large nano-spheres}
\label{sec:1}

Within the RPA in semiclassical limit \cite{jacak5}, 
 the solution of the dynamical equation for local density of electrons in a metallic nano-sphere with the radius $a$, can be decomposed into two parts  related   to  the distinct domains:
\begin{equation}
 \delta \tilde{\rho}( {\bmb r,t})=\left\{
           \begin{array}{l}
             \delta \tilde{\rho}_1( {\bmb r,t}), \;for\; r<a,\\
              \delta \tilde{\rho}_2( {\bmb r,t}), \;for\; r\geq a,\; ( r\rightarrow a+),\\
          \end{array}
       \right.
       \end{equation}
corresponding to the volume and surface  excitations, respectively. These two parts of local electron density fluctuations
satisfy the equations \cite{jacak5}:
\begin{equation}
\label{e20}
\frac{\partial^2 \delta \tilde{\rho}_1 ({\bmb r},t) }{\partial t^2}=\frac{2}{3} \frac{\epsilon_F}{m}\nabla^2 \delta \tilde{\rho}_1( {\bmb r},t)-
\omega_p^2 \delta \tilde{\rho}_1( {\bmb r},t),
\end{equation}
and
\begin{equation}
\label{e21}
\begin{array}{l}
\frac{\partial^2 \delta \tilde{\rho}_2 ({\bmb r},t) }{\partial t^2} =-
\frac{2}{3m} \nabla\left\{\left[\frac{3}{5}\epsilon_F n_e+\epsilon_F \delta \tilde{\rho}_2 
 {\bmb r},t)\right]\frac{\bmb r}{r}\delta\right\}\\
 -  \left[\frac{2}{3} \frac{\epsilon_F}{m}\frac{\bmb r}{r}\nabla \delta \tilde{\rho}_2( {\bmb r},t)
+      \frac{\omega_p^2}{4\pi}          \frac{\bmb r}{r}\nabla \int d^3r_1 \frac{1}{|{\bmb r}-{\bmb r}_1|}\right.\\
\left.\times \left(\delta \tilde{\rho}_1( {\bmb r}_1 ,t)
\Theta(a-r_1)  
+\delta \tilde{\rho}_2( {\bmb r}_1 ,t)\Theta(r_1-a)\right)\right]\delta,\\
\end{array}
\end{equation}
where,  $\omega_p^2=\frac{4\pi n_e e^2}{m}$  is the bulk plasmon frequency, 
$\Theta$ is the Heaviside step function, $\delta=\delta(r-a)$. The analysis and solutions of the above equations are performed in details as presented  in Ref.
\onlinecite{jacak5}, resulting in determination of plasmon self-mode spectrum, both for volume and surface modes. 

Nevertheless, this treatment did not account for plasmon attenuation. 
One  can, however,  include damping of plasmons in a  phenomenological manner,  adding an attenuation  term  to  plasmon dynamic equations, i.e., by adding the term,$-\frac{2}{\tau_0}\frac{\partial \delta\rho({\bmb r},t)}{\partial t}$, to the r.h.s.
  of both Eqs (\ref{e20}) and (\ref{e21}), taking advantage of their oscillatory form   \cite{jacak5}. Except of homogeneous equations (\ref{e20}) and (\ref{e21}) determining self-frequencies of plasmon modes, the dual inhomogeneous would be written, with explicit expression of forcing factor. This factor would be the time dependents electric field, including electrical component of e-m wave. 
For e-wave frequency in resonance with plasmons in the metalillic nanosphere,  the wave-length (being of order of 500 nm) highly exceeds  the nanosphere size (with radius $10-50$ nm), thus the dipole regime is in force. 
 For the homogeneous forcing field ${\bmb E}(t)$ (which corresponds  to dipole approximation satisfied for $a\sim 10-50$ nm, when $\lambda \sim 500$ nm),  only dipole surface mode can be excited  and the electron response 
resolves to  a single dipole type mode, described by the function $Q_{1m}(t)$.
 The  function $Q_{1m}(t)$ satisfies the equation:
  \begin{equation}
  \label{qqq}
  \begin{array}{l}
  \frac{\partial^2Q_{1m}(t)}{\partial t^2}+\frac{2}{\tau_0}\frac{\partial Q_{1m}(t)}{\partial t}+\omega_1^2 Q_{1m}(t)\\
   =\sqrt{\frac{4\pi}{3}}\frac{en_e}{m}\left[E_z(t)\delta_{m0}+\sqrt{2}\left(E_x(t)\delta_{m1}
   + E_y(t)\delta_{m-1}\right)\right],\\
   \end{array}
   \end{equation}
   where  $\omega_1=\omega_{01}=\frac{\omega_p}{\sqrt{3\varepsilon}}$ (it is a dipole-type surface plasmon Mie frequency \cite{Mie}).
   Only this function contributes to the plasmon response to the homogeneous electric field.
Thus for the homogeneous forcing field, electron density fluctuations \cite{jacak5}:
\begin{equation}
\label{oscyl}
 \delta \rho({\bmb r},t)=\left\{
  \begin{array}{l}
  0,\;\; r<a,\\
\sum\limits_{m=-1}^{1}Q_{1m}(t)Y_{1m}(\Omega)\; r\geq a,\; r\rightarrow a+.\\
  \end{array} \right.
  \end{equation}

For plasmon oscillations given by Eq. (\ref{oscyl}) one can calculate the corresponding dipole,
\begin{equation}
\label{dipolek}
{\bmb D}(t)= e\int d^3r {\bmb r}\delta\rho({\bmb r},t)=  \frac{4\pi}{3}e{\bmb q}(t)a^3,
\end{equation}
where,
$Q_{11}(t)=\sqrt{\frac{8\pi}{3}}q_x(t)$,  $Q_{1-1}(t)=\sqrt{\frac{8\pi}{3}}q_y(t)$,\\
   $Q_{10}(t)=\sqrt{\frac{4\pi}{3}}q_x(t)$
   and ${\bmb q}(t)$ satisfies the equation (cf. Eq. (\ref{qqq})),
   \begin{equation}
   \label{dipoleq}
   \left[\frac{\partial^2}{\partial t^2}+  \frac{2}{\tau_0}  \frac{\partial}{\partial t} +\omega_1^2\right] {\bmb q}(t)=\frac{en_e}{m}
   {\bmb E}(t).
   \end{equation}

There are various mechanisms of plasmon damping, which could be effectively accounted for via phenomenological
oscillator type damping term. All types of scattering phenomena, including electron-electron and electron-phonon interactions,
as well contribution of the  boundary scattering effect \cite{atwater}, cause significant attenuation of plasmons, in particular,
in small metal clusters. All these contributions to damping time ratio
scale as $\frac{1}{a}$ and are of lowering significance with the radius growth. In the following subsection we
argue that damping of plasmons caused by  radiation losses scales conversely, as $a^3$, and for large
nano-spheres this channel dominates plasmon attenuation.

\subsection{Lorentz friction for plasmons}

Plasmon oscillations
  are themselves a source of the e-m radiation. This radiation takes away the energy of plasmons resulting
  in their damping, which can be described as the Lorentz friction force reducing charge oscillations \cite{lan}. This damping was not included in $\tau_0$ in Eq. (\ref{dipoleq}). This $\tau_0$ accounted only for 
 scattering of electrons on other electrons, on defects, on
    phonons and on nanoparticle boundary---all they lead to damping rate expressed by the simplified formula \cite{atwater}:
\begin{equation}
\label{form}
\frac{1}{\tau_0}\simeq \frac{v_F}{2\lambda_b }+\frac{cv_F}{2a},
\end{equation}
where, $C$ is the constant of unity order,  $a$ is the nano-sphere radius, $v_F$ is the Fermi velocity in metal,
  $\lambda_b$ is the electron free path in bulk (including  scattering of electrons on other electrons,
  on impurities and on phonons \cite{atwater}); for  Ag, $v_F=1.4\times 10^6$ m/s and $\lambda_b\simeq 57$ nm
   (at room temperature); the latter term in the formula (\ref{form})  accounts for scattering of electrons on the boundary of
    the nanoparticle, while the former one corresponds to  scattering processes similar as in bulk. The other effects, as
     the so-called Landau damping (especially important in small clusters \cite{jo,ekardt2}), corresponding to decay of
      plasmon for  high energy  particle-hole pair, are   of lowering significance   for nano-sphere  radii larger than $ 2-3$ nm
       \cite{jo}  and  are completely negligible for radii larger than 10 nm. Note that the similarly lowering role with the
        radius growth plays also  electron liquid spill-out  effect \cite{brack,ekardt},  though it was of
         primary importance for small clusters \cite{brack,kresin}.

   The electron
   friction  caused  by e-m wave emission  can be described as the additional electric field \cite{lan},
\begin{equation}
\label{lorentz}
  {\bmb E}_L = \frac{2}{3\varepsilon^{3/2}v^3}\frac{\partial^3{\bmb D}(t)}{\partial t^3},
\end{equation}
where $v=\frac{c}{\sqrt{\varepsilon}}$ is the light velocity in the dielectric medium, and ${\bmb D}(t)$ is the dipole of the nano-sphere.
According to Eq. (\ref{dipolek}) we arrive at the following:
\begin{equation}
\label{lor}
{\bmb E}_L= \frac{2e}{3\varepsilon v^2}\frac{4\pi}{3}a^3\frac{\partial^3{\bmb q}(t)}{\partial t^3}.
\end{equation}
Substituting this into Eq. (\ref{dipoleq}), we get,
\begin{equation}
\begin{array}{l}
\left[\frac{\partial^2}{\partial t^2}+  \frac{2}{\tau_0}  \frac{\partial}{\partial t} +\omega_1^2\right] {\bmb q}(t)\\
=\frac{en_e}{m}
   {\bmb E}(t) +\frac{2}{3\omega_1}\left(\frac{\omega_1a}{v}\right)^3\frac{\partial^3{\bmb q}(t)}{\partial t^3}.
\end{array}
   \end{equation}
   If one rewrites the above equation (for ${\bmb E}$=0) in the form,
   \begin{equation}
   \label{appr1}
\left[\frac{\partial^2}{\partial t^2} +\omega_1^2\right] {\bmb q}(t)=
 \frac{\partial}{\partial t}\left[ -\frac{2}{\tau_0} {\bmb q}(t) +
\frac{2}{3\omega_1}\left(\frac{\omega_1a}{v}\right)^3\frac{\partial^2{\bmb q}(t)}{\partial t^2}\right],
   \end{equation}
 thus, one notes that the zeroth order approximation (neglecting attenuation) corresponds to the equation:
 \begin{equation}
 \label{appr}
\left[\frac{\partial^2}{\partial t^2} +\omega_1^2\right] {\bmb q}(t)= 0.
\end{equation}
In order to solve Eq. (\ref{appr1}) in the next step of perturbation iteration, one can substitute, in the r.h.s. of this equation,
 $\frac{\partial^2{\bmb q}(t)}{\partial t^2}$ by $-\omega_1^2 {\bmb q}(t) $ (acc. to Eq. (\ref{appr})).

Therefore, if one assumes  the above estimation, 
$ \frac{\partial^3{\bmb q}(t)}{\partial t^3}\simeq -\omega_1^2    \frac{\partial{\bmb q}(t)}{\partial t}$,
 one can include the Lorentz friction in a  renormalized damping term:
\begin{equation}
\label{ratio}
  \left[\frac{\partial^2}{\partial t^2}+  \frac{2}{\tau}  \frac{\partial}{\partial t} +\omega_1^2\right] {\bmb q}(t)=\frac{en_e}{m}
   {\bmb E}(t) ,
   \end{equation}
   where,
   \begin{equation}
   \label{tau}
   \frac{1}{\tau}=\frac{1}{\tau_0}+\frac{\omega_1}{3}\left(\frac{\omega_1 a}{v}\right)^3\simeq \frac{v_F}{2\lambda_B}+\frac{Cv_F}{2a}
   +  \frac{\omega_1}{3}\left(\frac{\omega_1 a}{v}\right)^3,
   \end{equation}
 and we used for $\frac{1}{\tau_0}\simeq  \frac{v_F}{2\lambda_B}+\frac{Cv_F}{2a} $
   \cite{atwater}.
The renormalized  damping causes a change in the shift of self-frequencies of free surface plasmons,
 $\omega_1'=\sqrt{\omega_1^2-\frac{1}{\tau^2}}$, which can be compared with the experimental observations for various nanosphere radii \cite{jacak11}.

Note also, that one can verify \cite{jacak11} the above calculated Lorentz friction contribution to plasmon damping by the
estimation of the energy
transfer in the far-field zone (which can be expressed by the Poynting vector) and via comparison with the energy loss of
plasmon oscillations. We have arrived \cite{jacak5,jacak11} at the same formula for damping time rate as given by  Eq. (\ref{ratio}).
The radius dependent shift of the resonance resulting due to strong irradiation-induced plasmon damping
was verified  experimentally \cite{jacak11} by measurement of light extinction in colloidal solutions of
nanoparticles with different size (it has been done \cite{jacak11} for Au, $10-80$ nm, and Ag, $10-60$ nm). These
measurements clearly support the $a^3$  plasmon damping scaling, as described above for the far-field zone radiation losses in a dielectric surroundings.

If, however, in the vicinity of the nano-sphere the another charged system is located, the situation would change. For instance, in the case
when the nano-sphere is deposited on the semiconductor surface, the near-field coupling of plasmons with semiconductor
band electrons must be included.

\section{Enhancement of energy transfer from plasmons to electric receiver located in the near-field zone}

Even if the derivation of plasmon dynamics equation  in the form of effective harmonic oscillator equation  is rigorous upon quantum approach of quasiclassical RPA method \cite{jacak5}, the inclusion of plasmon attenuation of scattering type and of radiation losses type needs some phenomenological assumptions. They resolve themselves to extension of quantum RPA harmonic oscillator formulation to the damped oscillator equation form with attenuation described by heuristically assumed damping rates. It has been proved \cite{jacak5,jacak15} that radiation losses, in the case of the free far-field zone radiation (i.e., in the case of vacuum or dielectric surroundings of metallic nano-sphere with oscillating plasmons), can be accounted for as the Lorentz friction force \cite{lan}, in the manner as described in the previous section. When in the near-field zone (closer than the wave length corresponding to plasmon frequency) the energy receiver (i.e., other  system of chargesv, like semiconductor with its band system or another metallic nano-sphere as in the chain)  is located  the irradiation losses are dominated  by energy transfer via this near-field zone coupling channel. Presence of the charged system of the energy receiver in the vicinity of e-m emitting nanosphere with plasmons, modifies a retarded e-m potential of the emitting system and this modifies the Lorentz friction formula, which had been derived, in the standard form, for the dielectric surroundings \cite{lan}. In particular, an enhancement of plasmon radiation losses in the case when the nanoparticles with dipole Mie surface plasmons (excited by incident external light) are deposited on the semiconductor surface, lies behind the observed PV efficiency growth in new generation of solar cells, metallically modified \cite{wzr3a,wzmocn1,wzr2,konk,wzmocn2,mof}. In this case, the related attenuation rate can be also estimated by application of the Fermi golden rule to the semiconductor inter-band  transitions induced by dipole near-field coupling with plasmons \cite{jacak5,jacak15}. As it was proved \cite{jacak5}, the resulting attenuation rate scales with nano-sphere radius, $a$, in different manner in comparison to  far-field radiation, and with some correction and renormalization  expressed in terms of the band system parameters \cite{jacak5}. One can expect the similar behavior  in the case of the near-field coupling between nano-spheres in the chain, but for the sake of effectiveness of modeling one can assume that related attenuation rate has the form as that for the standard  Lorentz friction renormalized only by some coefficient phenomenologically assumed in order to account for the modification of e-m potential by the receiver system presence. 

\section{Nonlinear corrections to Lorentz friction force}

Let us consider a metallic nano-sphere located (the center) in $\mathbf{R}_0$. The electric dipole of electrons (fluctuation of electron density beyond the uniform distribution compensated by positive jellium) equals to,
\begin{equation}
    \mathbf{D}(\mathbf{R}_0,t)=e\int_V\delta \rho (\mathbf{r},t)\mathbf{r}d^3r.
\end{equation} 
This dipole corresponds to surface plasmons of dipole type which oscillates with Mie frequency $\omega_1=\omega_p/\sqrt{3\varepsilon}$ \cite{jacak5}, where $\omega_p$ is bulk plasmon frequency, $\varepsilon$ is the dielectric constant of the surrounding medium. These plasmons are not everlasting excitations and are damped due to scattering phenomena with the damping rate, 
$\frac{1}{\tau_0}=\frac{v_F}{2a}+\frac{Cv_F}{\lambda_b}$. For large nano-spheres the much more effective mechanism of plasmon damping are, however, irradiation energy losses, which for the case of irradiation to far-field zone can be expressed by the Lorentz friction \cite{jacak11,lan}. Assuming that electrons in the nano-sphere have positions $\mathbf{r}_i$ and assuming static jellium, the dipole of the nano-sphere,
$\mathbf{D}(\mathbf{R}_0,t)=e\sum_{i=1}^{N_e}\mathbf{r}_i=eN_e\mathbf{r}_e(t)$, where 
$\mathbf{r}_e =\sum_{i=1}^{N_e}\mathbf{r}_i/N_e$ is the mass center of the electron system. In the case of dynamics, the velocity of the mass center equals to, $\mathbf{v}_e=\sum_{i=1}^{N_e}\mathbf{v}_i/N_e$. 

On the charge $eN_e$, located in the mass center $\mathbf{r}_e(t)$ acts a Lorentz friction force \cite{lan},
\begin{equation}
    \mathbf{f}_L=\frac{2}{3}(eN_e)^2\left[\frac{d^2\mathbf{u}}{ds^2}-\mathbf{u}
\left(\frac{dU_j}{ds}\right)^2\right],\;\;j=1,...,4,
\end{equation}
where,
$ds=cdt\sqrt{1-v_e^2/c^2}$,
 $$
U_j=\left\{ \begin{array}{l}
\mathbf{u}=\mathbf{v}_e/(c\sqrt{1-v_e^2/c^2})\\
u_4=i/\sqrt{1-v_e^2/c^2}\\
\end{array}
\right. ,\;\;
U_j^2=-1.
$$
Up to terms of order $v_e^2/c^2$ with respect to the main term, one can write the electric field equivalent to  the Lorentz friction force, 
\begin{equation}
\begin{array}{l}
    \mathbf{E}_L(t)=\frac{\mathbf{f}_L}{eN_e}
=\frac{2}{3}(eN_e)\frac{1}{c^3}
\left\{\frac{d^2\mathbf{v}_e}{dt^2}+\frac{1}{c^2}\left[\frac{3}{2}\frac{d^2\mathbf{v}_e}{dt^2}
v_e^2\right.\right.\\
\left.\left.+3\frac{d\mathbf{v}_e}{dt}\left(\mathbf{v}_e\cdot \frac{d\mathbf{v}_e}{dt}\right)+\mathbf{v}_e\left(\mathbf{v}_e\cdot \frac{d^2\mathbf{v}_e}{dt^2}\right)\right]\right\}.\\
\end{array}
\end{equation}
Next, using dimensionless variables, 
$t'=t\omega_1$, $\mathbf{R}(t')=\frac{\mathbf{r}_e(t)}{a}$, $\dot{\mathbf{R}}(t')=\frac{d\mathbf{r}_e(t)}{a\omega_1dt}=\frac{\mathbf{v}_e}{a\omega_1}$,
$\ddot{\mathbf{R}}(t')=\frac{d^2\mathbf{r}_e(t)}{a\omega_1^2dt^2}=\frac{d\mathbf{v}_e}{a\omega_1^2dt}$,
$\stackrel{...}{\mathbf{R}}(t')=\frac{d^2\mathbf{v}_e(t)}{a\omega_1^3dt^2} $,
(dots indicate  derivatives with respect to $t'$), one can write out the dynamical equation in a convenient form.

Taking into account that the dipole corresponding to surface plasmons,
\begin{equation}
\label{eq1}
\bmb{D}=eN_{e}a\bmb{R},
\end{equation}
satisfies equation of  oscillatory-type, one can write it in the form (incorporating also the Lorentz friction force),
\begin{equation}
\label{dipol}
\begin{array}{l}   \stackrel{..}{\mathbf{R}}+\stackrel{}{\mathbf{R}}+\frac{2}{\tau_0\omega_1}\stackrel{.}{\mathbf{R}}
=\frac{2}{3}\left(\frac{\omega_p a}{\sqrt{3\varepsilon}c}\right)^3\left\{\stackrel{...}{\mathbf{R}}\right.\\
\left.+\left(\frac{\omega_p a}{\sqrt{3\varepsilon}c}\right)^2\left[\frac{3}{2}\stackrel{...}{\mathbf{R}} (\stackrel{.}{\mathbf{R}}\cdot \stackrel{.}{\mathbf{R}})
+3\stackrel{..}{\mathbf{R}}(\stackrel{.}{\mathbf{R}}\cdot \stackrel{..}{\mathbf{R}})+
\stackrel{.}{\mathbf{R}}(\stackrel{.}{\mathbf{R}}\cdot \stackrel{...}{\mathbf{R}})\right]\right\},\\
\end{array}
\end{equation}
the terms on r.h.s. of the above equation describe the Lorentz friction including relativistic nonlinear corrections (in bracket) beyond the ordinary main linear term $\sim \stackrel{...}{\mathbf{R}}$, as given by  (\ref{lorentz}).

For the case when $\frac{1}{\tau_0\omega_1}\simeq\left(\frac{\omega_pa}{c\sqrt{3\varepsilon}}\right)^3\ll 1$ (well fulfilled for nano-spheres with radii $10-50$ nm, Au or Ag), one can apply perturbation method of solution, and in zero order perturbation assume $\ddot{\mathbf{R}}+\mathbf{R}=0$. In the next step of perturbation one can thus substitute $\stackrel{..}{\mathbf{R}}=-\stackrel{}{\mathbf{R}}$ and 
$\stackrel{...}{\mathbf{R}}=-\stackrel{.}{\mathbf{R}}$ in the r.h.s. of the Eq. (\ref{dipol}).

Let us consider first a single metallic nano-sphere with dipole type surface oscillations with the dipole $\mathbf{D}$.
In the framework of the perturbation method of solution of dynamical equation of oscillatory type for the dipole, Eq. (\ref{dipol}), in the first order of  perturbation, attains the following  form (including the damping of plasmons due to scattering with the rate $\frac{1}{\tau_0}$ and due to radiation losses accounting for the linear term of Lorentz friction, while the r.h.s. of the equation (\ref{eq4}) expresses  nonlinear corrections to Lorentz friction), 
\begin{equation}
\label{eq4}
\begin{array}{l}
\ddot{\bmb{R}}+\bmb{R}+\left[\frac{2}{\tau_0 \omega_{1}}+\frac{2}{3}\left(\frac{\omega_p a}{\sqrt{3\varepsilon}c}\right)^{3}\right]\dot{\bmb{R}}\\
=\frac{2}{3}\left(\frac{\omega_p a}{\sqrt{3\varepsilon}c}\right)^{5}\left\{-\frac{5}{2}\dot{\bmb{R}}
\left(\dot{\bmb{R}}\cdotp\dot{\bmb{R}}\right)+3\bmb{R}\left(\dot{\bmb{R}}\cdotp\bmb{R}\right)\right\}.\\
\end{array}
\end{equation}

The above nonlinear equation is complicated in mathematical sense and advanced methods of solutions must be applied, as described in \cite{mit}. According to the special asymptotic methods for solution of nonlinear differential equation (\ref{eq4}), one can find  the solution in the following form ($\bmb{R}=R\frac{\bmb{r}}{r}$),
\begin{equation}
\label{eq5}
\begin{array}{l}
R(t)=\frac{A_0 e^{-\frac{t}{\tau}}}{\sqrt{1+\frac{9}{8}\gamma {A_0}^2\left(1-e^{-\frac{2t}{\tau}}\right)}}\cos\left(\omega_1 t+\Theta_0\right),\\
\end{array}
\end{equation}
with 
$\frac{1}{\tau \omega_1}=\frac{1}{\tau_0 \omega_1}+\frac{1}{3}\left(\frac{\omega_p a}{\sqrt{3\varepsilon}c}\right)^3\approx\frac{1}{3}\left(\frac{\omega_p a}{\sqrt{3\varepsilon}c}\right)^3$ (what is satisfied for $a$ larger than 10 nm),
$\gamma=\tau \omega_1\frac{1}{3}\left(\frac{\omega_p a}{\sqrt{3\varepsilon}c}\right)^5.$

\begin{figure}
  \scalebox{1.0}{\includegraphics{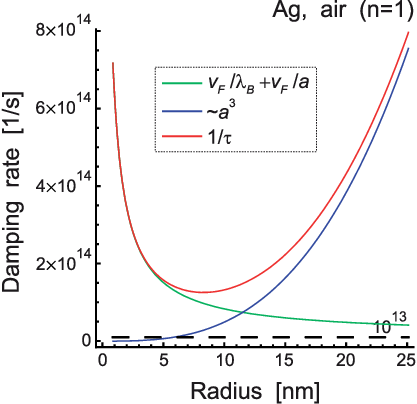}}
\caption{Contribution to the damping rate of surface plasmon oscillations in the nano-sphere versus the nano-sphere radius, including the scattering attenuation (green line) and the linear Lorentz friction damping (blue line); for radii greater than ca 10 nm the second channel dominates in overall damping (red line)}
\label{fig1}       
\end{figure}

In the formulae (\ref{eq5}) both coefficients $\frac{1}{\tau \omega_1}$ and $\gamma$ can be renormalized eventually by the mentioned above phenomenological factor accounting for the change of e-m potential caused by the presence of energy receivers in near-field zone of plasmons on the considered metallic nano-sphere (still holding here 1 for simplicity). From the form of equation (\ref{eq5}) it follows that $\frac{1}{\tau \omega_1}$ is always positive. Note that the scattering term, $\frac{1}{\tau_0}=\frac{v_F}{2a}+\frac{Cv_F}{2\lambda_b}$,  is negligible (for nano-sphere radius beyond 10 nm) in comparison with the linear contribution of the Lorentz friction, as it is demonstrated in  Fig. \ref{fig1}.

 The scale of the nonlinear corrections is given by the coefficient $\gamma\approx 10^{-4}\left(\frac{a[nm]}{10}\right)^2$. As this coefficient is small, one can neglect the related contribution in the denominator for the dipole solution (\ref{eq5}), which results in ordinary linear solution of damped oscillations. It means that the nonlinear corrections to the Lorentz friction have no significance in the case of plasmon oscillations of a single nano-sphere. This situation changes, however, considerably in the case of collective plasmon excitation propagating along the metallic nano-chain, as it will be described in the following paragraph.

\section{Collective plasmon wave-type propagation along the nano-chain in the nonlinear regime}

\begin{figure}
 \scalebox{0.42}{\includegraphics{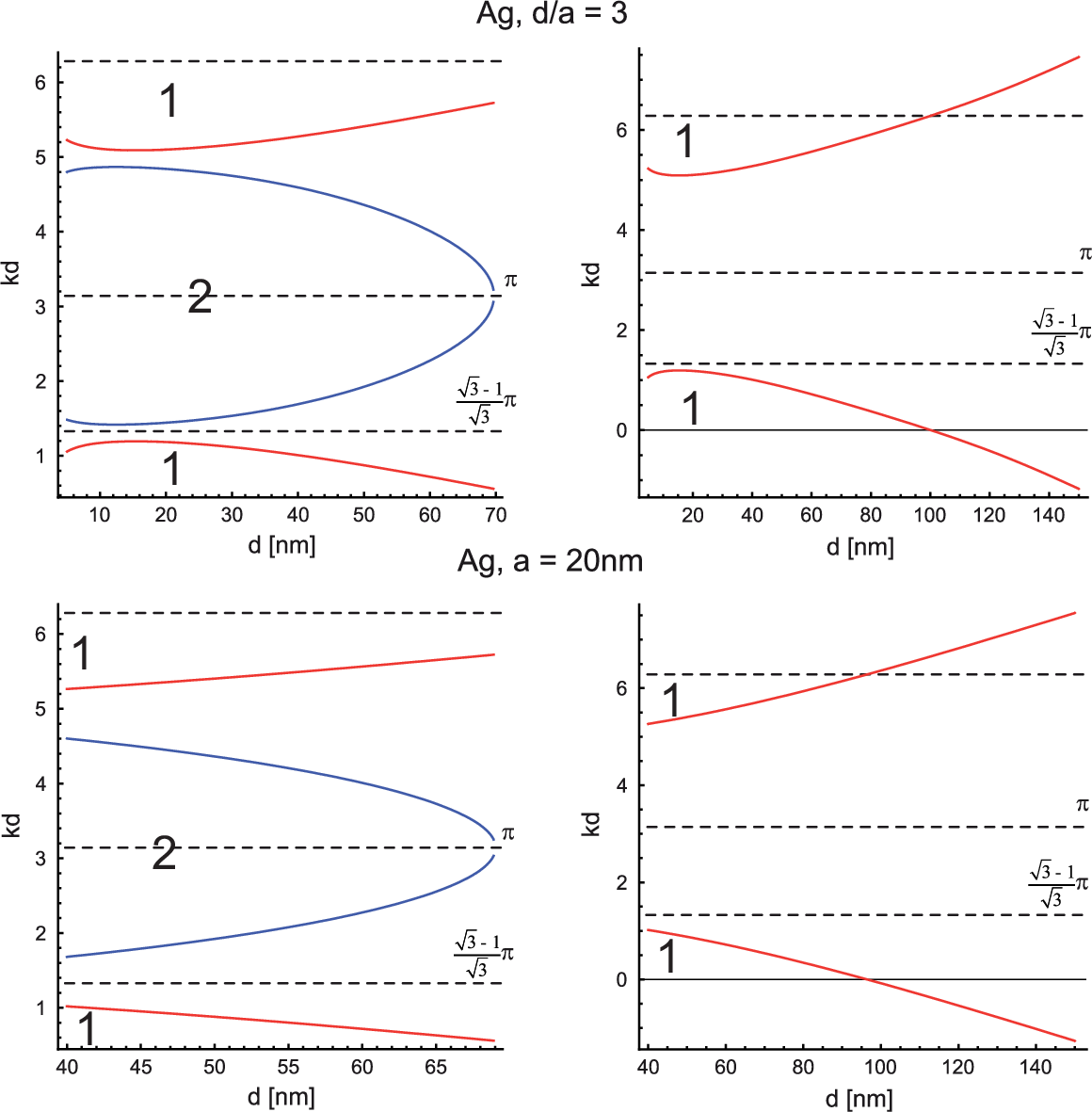}}
\caption{Regions for negative value of damping rates for plasmon-polaritons in the chain (1 for longitudinal polarization modes and 2 for transversal one,  in the nonlinear formulation framework; for the linear theory, red lines gives the position of vanishing damping rate for longitudinal modes of plasmon-polaritons and blue lines the same for transversal modes}
\label{fig2}      
\end{figure}

In the case of the metallic nano-chain one has to take into account the mutual affecting of nano-spheres in the chain. Assuming that in the sphere located in the point $\mathbf{r}$ we deal with the dipole $\mathbf{D}$, then in the other place $\mathbf{r}_0$ ($\mathbf{r}_0$ is fixed to the end of $\mathbf{r}$) the dipole type electric field attains the form as follows (including electro-magnetic retardation),
\begin{equation}
\label{eq6}
\begin{array}{l}
\bmb{E}\left(\bmb{r},\bmb{r}_0,t\right)=\frac{1}{\epsilon {r_0}^3}\left\{3\bmb{n}_0\left(\bmb{n}_0\cdotp\bmb{D}\left(\bmb{r},t-\frac{r_0}{v}\right)\right)\right.\\
\left.-\bmb{D}
\left(\bmb{r},t-\frac{r_0}{v}\right)\right\},\;\;
\bmb{n}_0=\frac{\bmb{r}_0}{r_0}, v=\frac{c}{\sqrt{\epsilon}}.
\end{array}
\end{equation}
This allows for writing out the dynamical equation for plasmon oscillations at each nano-sphere of the chain, which can be numbered by integer $l$ ($d$ will denote the separation between nano-spheres in the chain; vectors $\mathbf{r}$ and $\mathbf{r}_0$ are collinear, if the origin is associated with one of nano-spheres in the chain). The first term of the right-hand-side in the following formula (\ref{dipoll}) describes the dipole type coupling between nano-spheres \cite{jacak10} and the other two terms correspond to contribution due to plasmon attenuation (in the latter term the Lorentz friction caused electric field accounts also for nonlinear corrections). The index $\alpha$ enumerates polarizations, longitudinal and transversal ones with respect to the chain orientation.
\begin{equation}
\label{dipoll}
\begin{array}{l}
\ddot{R}_{\alpha}+R_{\alpha}\left(ld,t\right)
=\sigma_{\alpha}\frac{a^3}{d^3}
\sum\limits_{m=-\infty,m\neq l}^{\infty}\frac{R_{\alpha}\left(md,t-\frac{d|l-m|}{v}\right)}{|l-m|^3}\\
-\frac{2}{\tau_0 \omega_1}\dot{R}_{\alpha}\left(ld,t\right)+\frac{e}{ma{\omega_1}^2}E_{\alpha}\left(ld,t\right),\\
\end{array}
\end{equation}
where, $\sigma_{\alpha}=\left\{\begin{array}{l}-1,\alpha=x,y\\2,\alpha=z\end{array}\right.$ is introduced to distinguish both  polarizations.
The summation in the first term of the r.h.s. of the equation (\ref{dipoll}) can be explicitly performed in the manner as presented in \cite{jacak10}, because it 
is the same as for the linear theory formulation. Similarly as in the linear theory framework, one can change to the quasi-momentum  picture, taking advantage of the chain periodicity (in analogy to Bloch states in  crystals with the reciprocal lattice of quasi-momentum), i.e.,
\begin{equation}
\label{eq8}
\begin{array}{l}
R_{\alpha}\left(ld,t\right)=\tilde{R}_{\alpha}\left(k,t\right)e^{-ikld},\\
0\leq k \leq\frac{2\pi}{d},\tilde{R}_{\alpha}\left(k\right)\sin\left(t\omega_1+\beta\right).
\end{array}
\end{equation}
Thus the equation (\ref{dipoll}) can be rewritten in the following form (the Lorentz friction term was represented similarly as in equation (\ref{eq4})),
\begin{equation}
\label{eq9}
\begin{array}{l}
\ddot{\tilde{R}}_{\alpha}\left(k,t\right)+{\tilde{\omega}_{\alpha}}^2 \tilde{R}_{\alpha}\left(k,t\right)\left\{\frac{1}{\tau_{\alpha} \omega_1}\right.\\
\left.  +\frac{1}{3}\left(\frac{\omega_p}{\sqrt{3\varepsilon}c}\right)^5\left(\frac{5}{2}|
\dot{\tilde{R}}_{\alpha}\left(k,t\right)|^2-3|\tilde{R}_{\alpha}\left(k,t\right)|^2\right)\right\}=0,\\
\end{array}
\end{equation}
where,
${\tilde{\omega}_{\alpha}}^2 = 1 - 2\sigma_{\alpha}\frac{a^3}{d^3}\cos\left(kd\right)\cos\left(\frac{d\omega_1}{v}\right)$ and 
\begin{equation}
\label{tlumienie}
\begin{array}{l}
\frac{1}{\tau_{x,y}\omega_1}=\frac{1}{\tau_0\omega_1}+\frac{1}{4}\left(\frac{\omega_1 d}{v}\right)\frac{a^3}{d^3}\left(\left(\frac{\omega_1 d}{v}\right)^2-\left(kd-\pi \right)^2+\frac{\pi^2}{3}\right),\\
\frac{1}{\tau_{z}\omega_1}=\frac{1}{\tau_0\omega_1}+\frac{1}{2}\left(\frac{\omega_1 d}{v}\right)\frac{a^3}{d^3}\left(\left(\frac{\omega_1 d}{v}\right)^2-\left(kd-\pi \right)^2+\frac{\pi^2}{3}\right).\\
\end{array}
\end{equation}
In the above formulae the remarkable property is linked with the expressions for the attenuation rate for both polarizations. Two last expressions below  equation (\ref{eq9}) give these damping rates explicitly and one can notice that they could change their signs depending on  values for $d$, $a$ and $k$. In Fig. (\ref{fig2}) the regions of negative value for damping rates are  marked (for both polarizations).

Applying the same methods for solution of the nonlinear equation (\ref{eq9}) as in the former paragraph, using the asymptotic methods \cite{mit}, one can find the corresponding solutions for both regions with positive and negative damping rate, respectively.

For the positive damping rate, $ \frac{1}{\tau_{\alpha}\omega_1}>0$,
\begin{equation}
\label{eq10}
\begin{array}{l}
\tilde{R}_{\alpha}\left(k,t\right)=\frac{A_{\alpha 0}e^{-\frac{t}{\tau_{\alpha}}}}{\sqrt{1+\gamma_{\alpha}A_{\alpha 0}^2\left(1-e^{-\frac{2t}{\tau_{\alpha}}}\right)}}\cos\left(\omega_{\alpha}t+\Theta_0\right),\\
\tilde{R}_{\alpha}\left(k,t\right)\rightarrow_{(t\rightarrow \infty)} 0,\\
\end{array}
\end{equation}
where,
$\gamma_{\alpha}=|\tau_{\alpha}\omega_1|\left(\frac{\omega_1 a}{c}\right)^3\frac{1}{4}\left(\frac{5}{2}\tilde{\omega}_{\alpha}^2-1\right)$. We note from the form of Eq. (\ref{eq10}) that this is a damped mode, vanishing at longer time scale.

Nevertheless, for negative damping rate, $\frac{1}{\tau_{\alpha}\omega_1}<0$, 
the solution has a different form, 
\begin{equation}
\begin{array}{l}
\tilde{R}_{\alpha}\left(k,t\right)=\frac{A_{\alpha 0}e^{\frac{t}{|\tau_{\alpha}|}}}{\sqrt{1+\gamma_{\alpha}A_{\alpha 0}^2\left(e^{\frac{2t}{|\tau_{\alpha}|}}-1\right)}}\cos\left(\omega_{\alpha}t+\Theta_0\right)\\
\tilde{R}_{\alpha}\left(k,t\right)\rightarrow_{(t \rightarrow \infty)}\frac{1}{\sqrt{\gamma_{\alpha}}}\cos\left(t\omega_{\alpha}+\Theta_0\right).\\
\end{array}
\end{equation}
This solution is stable. It corresponds to an undamped mode which stabilizes on the fixed amplitude at longer time scale, independent of initial conditions expressed by $A_{\alpha}$.

The corresponding dipole oscillations attain
in the latter case the form of 'planar' waves propagating along the chain,
\begin{equation}
\label{eq14}
D_{\alpha}=\frac{e N_e a}{\sqrt{\gamma_{\alpha}}}\frac{1}{2}\left\{e^{i\left(\omega_{\alpha}t-kld\right)}+e^{-i\left(\omega_{\alpha}+kld\right)}\right\}.
\end{equation}

From the above formulae it follows that for positive attenuation rate we deal with ordinary damped plasmon-polariton propagation, not strongly modified in comparison to linear theory (due to small value of the factor $\gamma$). Nevertheless, in the case of negative damping rate the solution behaves differently--on longer time scale this solution stabilizes on the constant amplitude  independently of initial conditions. This remarkable property characterizes undamped propagation of plasmon-polariton along the chain. If one turns back to dipole explicit form (\ref{eq1}), then typical 'planar' wave formula with constant amplitude describes this undamped mode, as written in the  equation (\ref{eq14}). The region of negative damping correspond thus, within the nonlinear approach, to undamped modes with the fixed amplitude. Let us note that  the same region was linked with instability of the linear theory (which was, however, the artefact of the linear approach).

Finally, one can calculate the group velocity of the undamped plasmon-polariton mode, in the following form,
\begin{equation}
\label{eq11}
v_{\alpha}=\frac{d\omega_{\alpha}}{dk}=\omega_1 d\frac{\sigma_{\alpha}a^3\sin\left(kd\right)\cos\left(\frac{d\omega_1}{c}\right)}{d^3\sqrt{1-2\sigma_{\alpha}\frac{a^3}{d^3}\cos\left(kd\right)\cos\left(\frac{d\omega_1}{c}\right)}}.
\end{equation}
From this formula it follows that the group velocity of the unadamped wave type collective plasmon excitation (called  plasmon-polariton) may attain different values depending on $a$, $d$ and $k$.

Indicated above undamped mode of propagation of collective surface plasmons seems to match with experimentally observed log range propagation of plasmon excitations along the metallic nano-chain \cite{maradudin,atwater1,atwater,ggg,plasmons}. The constant and fixed value of the amplitude for these oscillations (\ref{eq10}) are independent of initial conditions, which means that these excitations will be present in the system even if are excited by arbitrary small fluctuations. Thus one can conclude that they are self-exciting modes which always present in the system provided that radii of spheres and their separation in the chain have values for which at least one of the attenuation rates (\ref{tlumienie}) is negative.

\section{Conclusions}
We have demonstrated the practical utilization of RPA semiclassical description of plasmon oscillations in metallic nano-spheres. The oscillatory form of dynamics both for volume and surface plasmons, rigorously described upon the RPA semiclassical limit fits well  with the large nano-sphere case, when the nano-sphere radius is greater than 10 nm and lower than  $60$ nm, (for Au, Ag or Cu material), what is confirmed by experimental observations, on the other hand. The most important property of plasmons on such large nano-spheres is the very strong  e-m irradiation caused by these excitations, which results in  quick damping of oscillations. The attenuation effects for plasmons were not, however, included into the quantum RPA model. Nevertheless, they could be included by a phanomenological manner, taking advantage of the oscillatory form of dynamical equations. Some information on plasmon damping can be taken from microscopic analyzes of small metallic clusters (especially made by LDA and TDLDA methods of numerical simulations employing Kohn-Sham equation). For larger nano-spheres, these effects, mainly  of scattering type (also Landau damping), are, however, not specially important as diminishing with radius growth, as $\frac{1}{a}$.

 The irradiation effects overwhelming the energy losses in the case of large nano-spheres can be grasped in terms of the Lorentz friction, which reduces the charge movement. This approach has been analyzed in the present paper. Two distinct situations were indicated, the first one--of the free radiation to far-field zone  in dielectric (or vacuum)  surroundings of single nanoparticle and the second one, when in the near-field zone of plasmons on the nano-sphere, an additional charged system is located. 

This additional charge system acting as the e-m energy receiver, strongly modifies the e-m potentials of the source and in this way modifies energy emission in comparison to the free emission in vacuum or in dielectric surroundings. In particular, the Lorentz friction is modified in the case of energy receiver presence in the near field zone of plasmonms,  in comparison to simple free emission to the far-field zone. The e-m energy receiver located close to emitting  nano-sphere, could be semiconductor (as in the case of metallically modified solar cells) or other metallic nano-spheres  (as in the case of metallic nano-chain). The latter situation has been analyzed  in this paper. We have shown previously \cite{jacak10} that along the infinite nano-chain  the collective plasmon-polaritons can propagate (being collective surface plasmons coupled by e-m field in near-field zone), which at certain values of nano-sphere radii and separation in the chain, appear as undamped modes. Simultaneously, the instability regions of linear theory of plasmon-polariton dynamics occur, which shows that the nonlinear corrections must be included. 

In this paper we have developed the nonlinear theory of collective plasmon-polariton dynamics along the chain, including nonlinear corrections to Lorentz friction force. Even though the related nonlinearity is small, it suffices to regularize the instable linear approach. As the most important observation, we noted the presence of undamped excitations (instead of those instable within the linear approach), which have fixed amplitude independently how small or large the initial conditions were. This excitations, typical for nonlinear systems, would have some practical significance, e.g., to enhance sensitivity of antennas with coverings by plasmon nano-systems offering self-induced collective plasmon-polaritons in wide range of frequencies, which would be excited by even very small signal (the energy to attain the stable level of plasmon-polariton amplitude would be supplied, in this case,  by an external auxiliary supply). 

\section{Acknowledgements}
Supported by Polish National Science Center project No: DEC-2011/03/D/ST3/02643

\bibliographystyle{spphys}       

\end{document}